\documentclass[journal]{IEEEtran}

\usepackage[dvipdfmx]{graphicx}
\usepackage[dvipdfm]{color}
\usepackage{amsmath,amssymb} 
\usepackage{multirow}
\usepackage[abs]{overpic}

\begin{document}

\title{High-contrast Coherent Population Trapping\\ based on Crossed Polarizers Method}

\author{Yuichiro~Yano 
        and~Shigeyoshi~Goka 
\thanks{Yuichiro~Yano is with the Department
of  and Electrical Engineering, Tokyo Metropolitan University, Minamiosawa, Hachioji,
Tokyo, 1-1 Japan e-mail: yano-yuichiro@ed.tmu.ac.jp}}


\maketitle

\begin{abstract}
A method based on crossed polarizers to observe high-contrast coherent population trapping (CPT) resonance has been developed.
Since crossed polarizers have a simple optical system, our method is suitable for chip-scale atomic clocks (CSACs).
In CPT, the Faraday rotation in a linearly polarized light field (lin$\parallel$lin) was calculated using two pairs of $\Lambda$-system models; the spectrum of the Faraday rotation is also estimated.
On measuring the contrast and linewidth with the crossed-polarizer method, a comparison of the theoretical model and experiment data showed good agreement.
A comparison of the theoretical model and experiment data showed they were in good agreement.
Moreover, the experimental results showed that a high contrast (88.4\%) and narrow linewidth (1.15 kHz) resonance could be observed using a Cs gas cell and D$_1$-line vertical-cavity surface-emitting laser (VCSEL).
\end{abstract}


\IEEEpeerreviewmaketitle

\section{Introduction}
Atomic clocks based on coherent population trapping (CPT) resonance with a vertical-cavity surface-emitting laser (VCSEL) have attracted attention as a means of fabricating very small atomic references, such as chip-scale atomic clocks (CSACs)\cite{Knappe}. CPT atomic clocks are in great demand for many applications, such as telecommunications, navigation systems, and synchronization of networks \cite{Vig}. Such atomic clocks are requrired for their high frequency stability, small volume and low power consumption. In particular, short-term frequency stability is an important parameter. Short-term frequency stability, described as the Allan standard deviation $\sigma_y(\tau)$, is estimated as 
\begin{equation}
\sigma _y (\tau)  \propto \frac{\Delta f}{f_0}\frac{1}{\mbox{\it SNR}}  \tau^{-1/2} 
\label{eq:short}
\end{equation}
\noindent
where $\Delta f$ is the resonance linewidth, $f_0$ is the resonance frequency, and $\mbox{\it SNR}$ is the signal-to-noise ratio. Contrast, which is used as a measure of $\mbox{\it SNR}$, is defined as the amplitude of CPT resonance over the background signal level \cite{Lutwak}.
Therefore, short-term stability is determined from contrast and linewidth.

A high-contrast CPT resonance can easily be obtained in a high-intensity laser field; however, the resonance linewidth broadens as a result of the power broadening effect {\cite{Powerbroadening}}.
It is difficult to both enhance contrast and sharpen the linewidth simultaneously.
To resolve this issue, a number of methods have been developed (e.g. double-lambda CPT {\cite{Pulsed CPT}}, push-pull pumping {\cite{Push-pull}}).
These methods increase the population of the clock transition through using the unique polarization of the incident laser beam to enhance the amplitude of CPT resonance signal.
For example, the double-lambda CPT is generated by exciting with two lin$\perp$lin polarized lasers.
And, push-pull pumping is a method to excite atoms using laser light of orthogonal linear polarizations with the time-delayed optical components.
These optical systems requires several optics (e.g. beam splitter, mirrors, retarders) to generate the unique polarization.
Therefore, because the optical system is complex and occupies a large volume,
it is difficult to use these methods for CPT atomic clocks requiring small volumes such as CSACs.

In contrast, dispersion detection methods have been reported to decrease the background signal level{\cite{Magneto}\cite{POP}}. 
This method regulates the light incident on a photo detector by selecting the light contributing to the resonance to reduce background signal.
Recently, the method was applied in the field of high sensitive magnetometry{\cite{Magneto}}.
And, it  was reported that by decreasing the background signal level it improved the Allan deviation by one order of magnitude in comparison with absorption detection employing pulsed optical pumped atomic clocks{\cite{POP}}.
The dispersion signal can be obtained by simply placing a polarizer or retarder in front of a photo detector along the same axis.
As the dispersion detection is a simple optical system,
it enables enhancements in the frequency stability of small volume CPT atomic clocks.

In this paper, we focus on the Faraday effect in CPT and propose a method based on crossed polarizers for observing high-contrast CPT.
Firstly, we calculate the Faraday rotation in CPT under a linearly polarized light field (lin$\parallel$lin) using the two pairs of a $\Lambda$ system model. The CPT lineshape can be estimated by calculating the Faraday rotation angle. Both the resonance amplitude and linewidth behavior of the magnetic field are estimated. Secondly, we show the results of an experiment using a $^{133}$Cs gas cell and the D$_1$-line VCSEL. The contrast and linewidth were measured by varying the magnetic field and the relative transmission angle of the polarizers. A comparison of the experimental and calculated results showed that they had the same tendency, and the difference between experiment and calculation was no more than 20 \%. Finally, on the basis of the measurement and calculation data, we determine the optimal condition for enhancing the short-term stability of CPT atomic clocks.

\begin{figure}[t]
\centering
\includegraphics{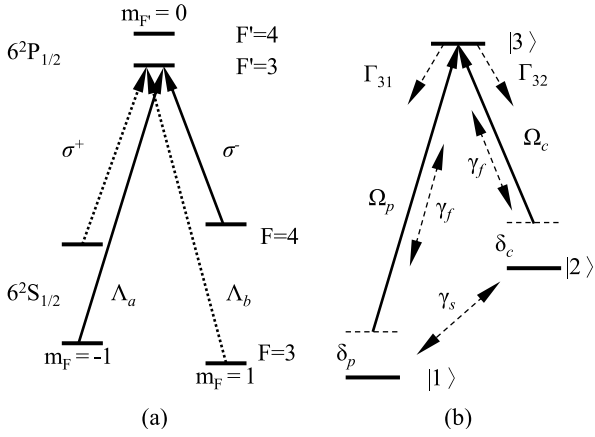}
\caption{(a) Excitation scheme with a lin$\parallel$lin field on the $\rm D_1$-line of Cs: the solid line is labeled $\Lambda _a$, and the dotted line is labeled $\Lambda_b$ (b) Closed $\Lambda$-type three-level model used to calculate the CPT phenomenon: $\delta_p$ and $\delta_c$ are detunings from the ground states, $\Omega_p$ and $\Omega_c$ are Rabi frequencies, $\Gamma_{31}$ and $\Gamma_{32}$ are relaxations between an excited state and the two ground states, and $\gamma_s$ is the relaxation term of the ground states. $\gamma_f$ is decoherence rate.}
\label{fig:scheme}
\end{figure}

\section{Theory}

\subsection{Magneto optical effect under CPT resonance}

\begin{figure}[t]
\centering
\begin{overpic}[scale=0.5]{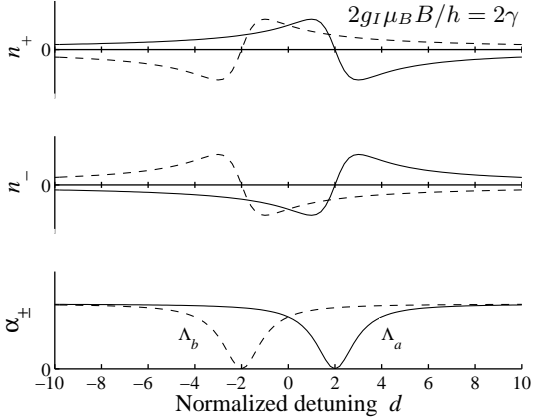}
\put(135,150){{\small $2g_I\mu_B B/h = 2\gamma$}}
\end{overpic}
\caption{Spectral dependences of absorption $\alpha$ and refractive index ($n_+$,$n_-$): the solid line corresponds to $\Lambda_a$ and the dash line to $\Lambda_b$, the magnitude of splitting is equal to twice the resonance line width $\gamma$.
}
\label{fig:spectrum_abs_ref}
\end{figure}

Faraday rotation is a magneto-optical phenomenon, that is, an interaction between light and a magnetic field in a medium. The resonant Faraday rotation is known as the Macaluso-Corbino effect \cite{Budker}. Though the Faraday rotation in CPT is classified as such an effect, the way to calculate it in CPT has not been reported yet. In this section, we describe the Faraday rotation in CPT in a lin$\parallel$lin light field.

Figure \ref{fig:scheme} (a) shows the excitation scheme using a lin$\parallel$lin light field on the $^{133}$Cs-D$_1$ line.
In CPT phenomenon under lin$\parallel$lin light field, two schemes can be formed with two pairs of ground-state hyperfine sublevels simultaneously:
$|F=3,m_F=1\rangle$,$|F=4,m_F=-1\rangle$ indicated by the label $\Lambda_a$ ,and $|F=3,m_F=-1\rangle$,$|F=4,m_F=1\rangle$ indicated label $\Lambda_b$ coupled with the common excited states $|F'=3,m_F=0\rangle$ or $|F'=4,m_F=0\rangle$ \cite{Zibrov}.
Therefore, four wave waves that result from a combination of two $\Lambda$ scheme and circularly polarization contribute the CPT phenomena under lin$\parallel$lin light field.

Figure \ref{fig:scheme} (b) shows a simple $\Lambda$-type three level system of the $\Lambda_a$ scheme.
Here, the energy eigenstates $|1\rangle$ and $|2\rangle$ correspond to two ground states $|F=3,m=-1\rangle $ and $|F=4,m=1\rangle$, and the excited state $|3\rangle$ correspond to $|F'=3,m=0\rangle$ or $|F'=4,m=0\rangle$.
Assuming equal Rabi frequencies $\Omega = \Omega_p = \Omega_c$ and decay rates $\Gamma_{31} =\Gamma_{32}$, the line shape for CPT resonance is Lorentzian.

In the presence of a magnetic field (C axis direction), the two resonances shift in opposite frequency directions because of the Zeeman effect.
The Zeeman shift of the two pairs in a weak magnetic field ($<$ 50 mT) is as follows:

\begin{equation}
f_{a,b}  = \\
f_0 \pm \\
\frac{2g_{I}^{}\mu_{B}^{}}{h}B+\frac{15g_{J}^{2}\mu_{B}^{2}}{32f_{0}h^2}B^2
\label{eq:zeeman-shift}
\end{equation}
\noindent
where $B$ is the magnetic field, and $f_0$ is the hyperfine splitting frequency of the ground states in the absence of the magnetic field, $g_I$ is the nuclear $g$-factor, $g_J$ is the Land\'e $g$-factor, $\mu_B$ is the Bohr magneton, and $h$ is Planck's constant.


\begin{table}
\renewcommand{\arraystretch}{1.3}
\caption{Absorption and refractive index}
\label{tab:abs_and_ref}
\centering
\begin{tabular}{c|c||c|c}
	\hline
	$\Lambda$& $\sigma$ &Absorption index  $\alpha$&Refractive index $n$\\
    \hline
    \multirow{2}{*}{$\Lambda_a$\rule[-3mm]{0mm}{10mm}} &
    $\sigma^+$ &
    $\chi_0\Bigl(1-\cfrac{\gamma^2}{(\delta-\frac{2g_I\mu_B}{\hbar}B)^2+\gamma^2}\Bigr)$&
	$-\cfrac{\chi_0\gamma(\delta-\frac{2g_I\mu_B}{\hbar}B)}{(\delta-\frac{2g_I\mu_B}{\hbar}B)^2+\gamma^2}$\rule[-4mm]{0mm}{10mm}\\ 
    \cline{2-4}&
    $\sigma^-$ &
	$\chi_0\Bigl(1-\cfrac{\gamma^2}{(\delta-\frac{2g_I\mu_B}{\hbar}B)^2+\gamma^2}\Bigr)$&
    $~~\cfrac{\chi_0\gamma(\delta-\frac{2g_I\mu_B}{\hbar}B)}{(\delta-\frac{2g_I\mu_B}{\hbar}B)^2+\gamma^2}$\rule[-4mm]{0mm}{10mm}\\ 
 
    \hline
    \multirow{2}{*}{$\Lambda_b$\rule[-3mm]{0mm}{10mm}} &
    $\sigma^+$ &
	$\chi_0\Bigl(1-\cfrac{\gamma^2}{(\delta+\frac{2g_I\mu_B}{\hbar}B)^2+\gamma^2}\Bigr)$&
    $~~\cfrac{\chi_0\gamma(\delta+\frac{2g_I\mu_B}{\hbar}B)}{(\delta+\frac{2g_I\mu_B}{\hbar}B)^2+\gamma^2}$\rule[-4mm]{0mm}{10mm}\\ 
    \cline{2-4}
    &$\sigma^-$ &
    $\chi_0\Bigl(1-\cfrac{\gamma^2}{(\delta+\frac{2g_I\mu_B}{\hbar}B)^2+\gamma^2}\Bigr)$&
    $-\cfrac{\chi_0\gamma(\delta+\frac{2g_I\mu_B}{\hbar}B)}{(\delta+\frac{2g_I\mu_B}{\hbar}B)^2+\gamma^2}$\rule[-4mm]{0mm}{10mm}\\ 
    \hline

\end{tabular}
\end{table}
The absorption index $\alpha$ and refractive index $n_+,n_-$ of the four light waves as a function of the normalized frequency detuning $d$ (= $\delta/\gamma$) is shown in Fig. {\ref{fig:spectrum_abs_ref}}.
All functions of absorption and refractive indices are shown in Table {\ref{tab:abs_and_ref}},
where $\chi_0$ represents the amplitude of linear susceptibility, $\delta$ the frequency detuning of ground state, and $\gamma$ the resonance linewidth (at half width at half maximum).

The absorption index is zero at the center of the CPT resonance
because the atoms do not interact with the light when the atoms fall into the dark state.
The absorptions $\sigma^-$ and $\sigma^+$ vanish simultaneously under CPT resonance.
Therefore, the absorption of left-circular-polarized light $\alpha_+$ equals that of right-circular-polarized light $\alpha_-$.

From the Kramers-Kronig relation, the refractive index is an odd function of frequency detuning because the absorption index is even function.
If $\delta_p$ and $\delta_c$ are both much less than $\gamma_f$, we obtain $\delta_p=-\delta_c =\delta/2$.
Because $\delta_p$ and $\delta_c$ have opposite signs and the dispersion spectra are odd functions of the detuning $\delta$, the refractive index $n_+$ and $n_-$ have opposite signs ($n_+=-n_-$).
The refractive indices for the two schemes $n_{\Lambda_a}$ and $n_{\Lambda_b}$ also have opposite signs ($n_{\Lambda_a}=-n_{\Lambda_b}$), because the direction of the circular-polarized light $\sigma^+$ and $\sigma^-$ is changed in the $\Lambda_a$ and $\Lambda_b$ schemes.

\subsection{Crossed Polarizers method}

\begin{figure}[t]
\centering
\includegraphics[width=3in]{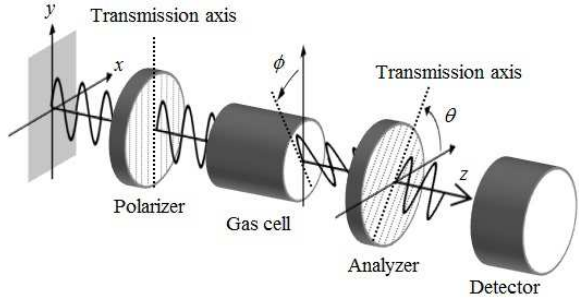}
\caption{Schematic layout of crossed polarizers method.}
\label{fig:schematic-cp}
\end{figure}

The short-term stability of laser-pumped vapor cell atomic clocks is often limited by a combination of light source AM noise and FM-AM conversion noise on the atomic absorption \cite{Kitching}.
The AM noise is caused by the power fluctuation of the light source. And FM-AM conversion noise is caused by the effect that laser frequency fluctuations have on the absorption \cite{Kitching}.
It is known that these noises are proportional to the background signal level (DC level) of the photo detector \cite{Lutwak}.
Therefore, to enhance the short-term stability of CPT atomic clocks, it is necessary to reduce the DC level.

The crossed polarizers method is a way of measuring the birefringence of the optical medium.
It has high sensitivity for birefringence detection because it suppresses the background signal level \cite{Dick}.
The schematic optical layout is shown in Fig. \ref{fig:schematic-cp}.
Two linear polarizers are placed on both sides of the gas cell, and the transmission axes are set nearly orthogonal to each other. The first polarizer and the second polarizer are called the polarizer and analyzer, respectively.
The transmission axes between the polarizer and analyzer are defined by the relative angle $\theta$.
In this paper, $\theta$ is defined as zero when the polarizers are orthogonal to each other.

Here, let the Faraday rotation angle induced by the gas cell be denoted $\phi$; the transmitted light $I_s$ can be written as 

\begin{equation}
\begin{aligned}
\frac{I_s}{I_0}=&\frac{1}{4}(e^{-2\pi\alpha_+ l/\lambda}-e^{-2\pi\alpha_- l/\lambda})^2\\
&+e^{-2\pi(\alpha_+ +\alpha_-)  l /\lambda}\sin^2 (\theta + \phi )\\
& \because \phi = \pi(n_+ -n_-)\frac{l}{\lambda}\\
\end{aligned}
\label{eq:dichroism_birefringence}
\end{equation}\noindent
where $I_0$ is the incident light intensity, $l$ the sample length, and $\lambda$ wavelength of light{\cite{Budker2}}.
In general, the two terms in Eq. ({\ref{eq:dichroism_birefringence}}) give comparable contributions to the forward-scattering signal.

The first term refers to the differential absorption of the $\sigma^+$ and $\sigma^-$ components (circular dichroism) and the second term to the differential dispersion.
Because $\alpha_+$ equals $\alpha_-$ at CPT resonance under lin$\parallel$lin field,
there is no difference in the absorption between $\sigma^+$ and $\sigma^-$.
Therefore, the dichroism term vanishes at CPT resonance.
Moreover, because the absorption coefficient $\chi_0 l / \lambda$ is small, taking into account that the conventional resonance contrast is no more than 10\% {\cite{contrast}},
we can assume that $e^{-2\pi(\alpha_+ +\alpha_-) l /\lambda}\approx 1$.
Eq.({\ref{eq:dichroism_birefringence}}) can be rewritten
\begin{equation}
\begin{aligned}
I_s=I_0\sin^2 (\theta + \phi ).
\end{aligned}
\end{equation}

The refractive index under CPT resonance is the same as the resonant Faraday effect\cite{Budker}.
The Faraday rotation angle $\phi$ is wrriten as 

\begin{equation}
\begin{aligned}
&\phi=\pi(n_+ - n_-)\frac{l}{\lambda}\\
&= \frac{\pi\chi_0l}{\lambda}\Bigl(\frac{(\delta-\frac{2g_I\mu_B B}{\hbar})\gamma}{(\delta-\frac{2g_I\mu_B B}{\hbar})^2+\gamma^2}-\frac{(\delta+\frac{2g_I\mu_B B}{\hbar})\gamma}{(\delta+\frac{2g_I\mu_B B}{\hbar})^2+\gamma^2}\Bigr),
\end{aligned}
\label{eq:faraday_rotation}
\end{equation}
\noindent
and by using a normalized detuning $d \equiv \delta/\gamma$ and normalized Zeeman shift $b \equiv (2g_I\mu_B B/\hbar)/\gamma$, Eq.(\ref{eq:faraday_rotation}) can be simplifiled into

\begin{equation}
\phi = \frac{\pi\chi_0\l}{\lambda}\frac{2 b(1+b^2-d^2)}{(1+b^2+d^2)^2-4b^2d^2}.
\label{eq:n-faraday-cpt}
\end{equation}

\subsection{CPT resonance characteristics with crossed polarizers method}

The polarization of wavelength components contributing CPT is rotated while passing through the cell.
On the other hand, the polarization of wavelength components not contributing CPT is not rotated.
Therefore, total transmitted light intensity $I$ can be expressed as

\begin{equation}
\begin{aligned}
I&=I_s + I_{nc} \sin^2(\theta)\\
&=I_c\sin^2 (\theta + \phi)+I_{nc} \sin^2(\theta)		
\end{aligned}
\label{eq:i_trans}
\end{equation}
\noindent
where $I_c$ is the sum of light intensities contributing to CPT, and $I_{nc}$ is the sum of light intensities not contributing to CPT.

Taking into account that the conventional resonance contrast is no more than 10\%,
the relation between $I_c$ and $I_{nc}$ is $I_c < I_{nc}$. Therefore, the DC level is dominated by the relative angle $\theta$.

When the relative angle $\theta$ is set larger than the phase shift $\phi$, which enables us to ignore the phase shift $\phi$, the DC level $I_{dc}$ is given as

\begin{equation}
I_\mathrm{\it dc} \approx I_c \sin^2(\theta) + I_{nc} \sin^2(\theta).
\label{eq:i_dc}
\end{equation}

And the resonance amplitude $I_s$ is 

\begin{equation}
I_s \approx \frac{\partial I_c}{\partial \theta} \phi = I_c \sin(2\theta)\phi
\label{eq:i_s}
\end{equation}

From Eq. (\ref{eq:i_s}), as the resonance amplitude is maximized by setting the relative angle $\theta$ to $\pi/4$. However, since the DC level increases with $\theta$ (from Eq. (\ref{eq:i_dc})), the highest contrast resonance can be obtained around $\theta \approx 0$.

When the relative angle $\theta$ is close to zero, the approximation that leads to Eq. (\ref{eq:i_s}) cannot be made because it relies on theta being much larger than $\phi$.
 Instead, from Eq. (\ref{eq:i_trans}) and using a small rotation approximation, the resonance amplitude $I_s$ is given as

\begin{equation}
I_s \simeq I_c \phi^2.
\label{eq:i_s2}
\end{equation}

\begin{figure}[]
\centering
\includegraphics[scale=0.5]{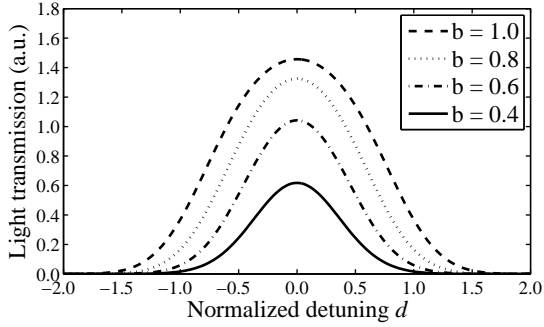}
\caption{Line shape of resonance with crossed polarizers calculated using Eq. (\ref{eq:n-faraday-cpt}) and Eq. (\ref{eq:i_s2}).}
\label{fig:lineshape-crossed polarizer}
\end{figure}

Next, we calculate the spectrum characteristics of the transmitted light. Figure \ref{fig:lineshape-crossed polarizer} shows the resonance lineshape calculated from Eq. (\ref{eq:n-faraday-cpt}) and Eq. (\ref{eq:i_s2}) when $\theta$ is set so as to minimize the DC level.
It is clear that the both the resonance amplitude and linewidth increase with increasing normalized magnetic field $b$.

The resonance amplitude $I_{pp}$, which is defined as the maximum change in the $I_s$ as a function of detuning, can be gotten from Eq. (\ref{eq:n-faraday-cpt}) and Eq. (\ref{eq:i_s2}):

\begin{equation}
I_{pp} =   I_c\frac{\pi^2 l^2\chi_0^2}{\lambda^2}\Big(\frac{(b^2+\sqrt{b^2+1}-1)(\sqrt{b^2+1}+2)}{2b(b^2+1)}\Big)^2
\label{eq:i_pp}
\end{equation}

\begin{figure}[]
\centering
\includegraphics[scale=0.5]{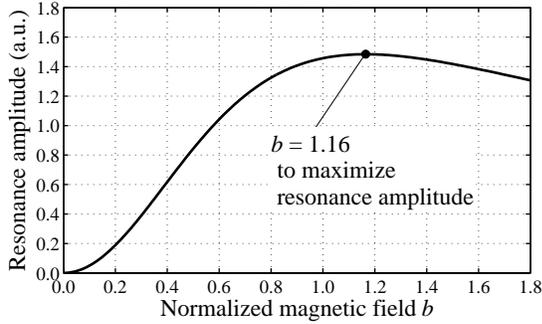}
\caption{Resonance amplitude as a function of normalized magnetic field $b$ calculated using Eq. (\ref{eq:i_pp}).}
\label{fig:signal-magnetic}
\end{figure}

Figure \ref{fig:signal-magnetic} plots the resonance amplitude as a function of the magnetic field using Eq. (\ref{eq:i_pp}). In the weak magnetic field ($b\ \leq 0.2$), the resonance amplitude is small because the Faraday rotation is small. In the range of $0.2< b \leq 1.16$, the resonance amplitude significantly increases with increasing magnetic field. The maximum resonance amplitude is obtained at $b= 1.16$. For magnetic fields $b$ over 1.16, the resonance amplitude tends to decrease because the two resonances separate from each other.

\begin{figure}[]
\centering
\includegraphics[scale=0.5]{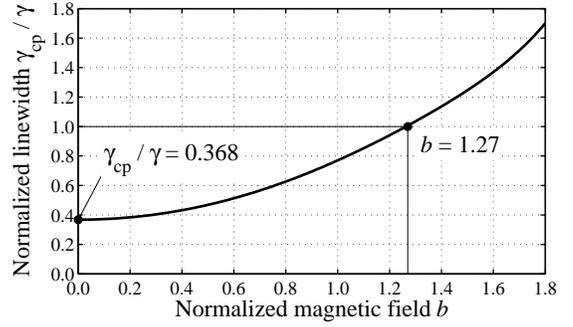}
\caption{Linewidth as a function of normalized magnetic field $b$ calculated using Eq. (\ref{eq:gamma_cp}).}
\label{fig:linewidth-magnetic}
\end{figure}

\begin{table}
\renewcommand{\arraystretch}{1.3}
\caption{$a_n$ values}
\label{tab:an}
\centering
\begin{tabular}{c||c|c|c|c}
    \hline
    $a_n$  &  $a_0$\ & $a_1$ & $a_2$ & $a_3$ \\
    \hline
Value & 3.681$\cdot$10$^{-1}$ & 3.988$\cdot$10$^{-1}$ & 1.644$\cdot$10$^{-2}$ & 3.706$\cdot$10$^{-4}$ \\
\hline
\end{tabular}
\end{table}

Figure \ref{fig:linewidth-magnetic} shows the linewidth with the crossed polarizers method $\gamma_{cp}$ normalized by the conventional linewidth $\gamma$ as a function of normalized magnetic field strength. The polynomial approximation of the normalized linewidth $\gamma_{cp}/\gamma $ is 

\begin{equation}
\gamma_{cp}/\gamma \approx \sum_{n=0}a_n b^{2n}
\label{eq:gamma_cp}
\end{equation}

where $a_n$ are constant values shown in Table \ref{tab:an} in the $n$ range from 0 to 3. The linewidth with the proposed method increases with increasing magnetic field. Moreover, since the linewidth is the sum of even functions and has a positive second derivative, the linewidth has a minimum value in the absence of the magnetic field. The minimum linewidth is the conventional one of 36.8\%. The linewidth of the proposed method is equal to that of the conventional one at a $b$ of 1.27.

From Fig. (\ref{fig:signal-magnetic}) and Fig. (\ref{fig:linewidth-magnetic}), we can obtain the relation between $b|_{I_{pp}=\max}$ and $b|_{\gamma_{cp}/\gamma=1}$:

\begin{equation}
b|_{I_{pp}=\max}<b|_{\gamma_{cp}/\gamma=1}
\label{eq:b-inequality}
\end{equation}

Thus, the linewidth of the proposed method is narrower than that with the conventional method in the magnetic field range from 0 to $b|_{I_{pp}=\max}$.

\section{Experimental setup}
\begin{figure}[t]
\centering
\includegraphics[scale=1]{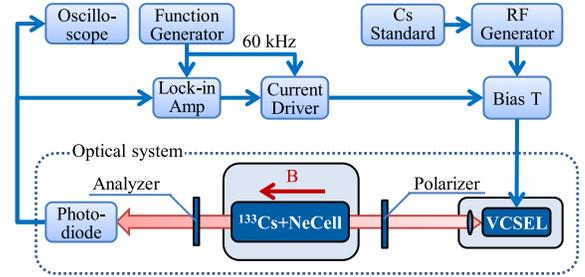}
\caption{Schematic diagram of experimental setup.}
\label{fig:setup}
\end{figure}
\begin{figure}[t]
\begin{center}
\includegraphics[scale=0.5]{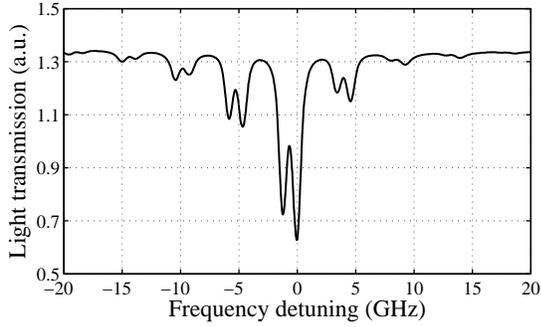}
\caption{Absorption profile of Cs-D$_1$ line using VCSEL modulated at 4.6 GHz: center frequency 335.1 THz (= 894.6 nm).}
\label{fig:absorption}
\end{center}
\end{figure}

Figure \ref{fig:setup} shows the experimental setup of the proposed observation method. The two polarizers were near-infrared sheet polarizers. A parallel linear beam (lin$\parallel$lin light field) was incident on the gas cell. The analyzer selected the optical polarization of wavelength components incident on the photodiode. The photodiode signal was connected to the oscilloscope.

A single-mode VCSEL fabricated by Ricoh Company, Ltd was used as the laser source.
The 894.6 nm wavelength of the VCSEL excites $^{133}$Cs at the D$_1$-line.
The VCSEL was driven by a DC injection current using a bias T and was modulated at 4.6 GHz using an analog signal generator to generate the first-order sidebands around the laser carrier. The absorption profile of the Cs-D$_1$ line using the VCSEL modulated at 4.6 GHz is shown in Fig. \ref{fig:absorption}. Since the incident light contains first-order sidebands and high-order sidebands, the plot shows some of the absorption lines. The two center absorption lines are excited by the first-order sidebands. Moreover, the minimum and second minimum peak correspond to two excited levels: $|F'=4\rangle$ and $|F'=3\rangle$. The frequency difference between the two excited states is wide enough that we can select either $|F'=3\rangle$ or $|F'=4\rangle$ as the absorption line. In this experiment, we selected $|F'=3\rangle$ as the excited state to stabilize the wavelength of the VCSEL.
	
A Pyrex gas cell containing a mixture of $^{133}$Cs atoms and Ne buffer gas at a pressure of 4.0 kPa was used. The gas cell was cylindrical, it had a diameter of 20 mm and optical length of 22.5 mm. The gas cell temperature was maintained at 42.0 $^\circ$C. The gas cell and Helmholz coil were covered with a magnetic shield to prevent an external magnetic field from affecting the magnetic field inside the cell. The internal magnetic field of the gas cell was created by the Helmholz coil. The internal magnetic field was calibrated using the frequency difference between magnetic-field sensitive and insensitive transitions by using $\rm \sigma$-$\rm \sigma$ excitation. The axis of the magnetic field was set to be parallel to the direction of the laser light (C-axis direction).

\section{Experimental Results and Discussion}
\subsection{Line shape of CPT resonance}

\begin{figure}[!b]
\centering
\includegraphics[scale=0.5]{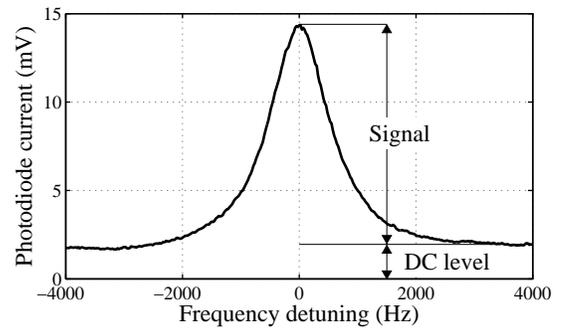}
\caption{Line shape of CPT resonance with crossed polarizers method. The line shape was obtained from one average scan. The applied static magnetic field was 93 $\mu$T (0.48 in normalized magnetic field). The incident light intensity was 1.1 mW/cm$^2$. The conventional resonance contrast and linewidth gotten by using a neutral density filter in place of the analyzer were respectively 3.3\% and 2.15 kHz under a static magnetic field of 5.0 $\mu$T.}
\label{fig:lineshape}
\end{figure}

Figure \ref{fig:lineshape} shows the observed CPT resonance with the crossed polarizers method. A good reduction in DC level was achieved because the transmission axis of the analyzer was optimized. Considering the transmitted light intensity, we estimate that the peak rotation angle is few tens of milli-radians. Since the signal was greater than the DC level, the conventional contrast, which was simply defined as the signal over the DC level, exceeded 100 \%. In this paper, contrast is defined so as not to exceed 100\% as follows.
\begin{equation}
\mbox{Contrast} := \frac{\mbox{Signal}}{\mbox{Signal} + \mbox{DC level}}
\label{eq:fhfs}
\end{equation}

Although the DC level was suppressed with the crossed polarizers method, weak light leakage was picked up by the photo detector. Since the leakage could not be reduced below the DC level by varying the relative angle $\theta$, the leakage was dependent on the extinction ratios of the polarizers. Owing to the DC level reduction, the proposed method yielded a contrast of 88.4\% with under a static magnetic field of 93 $\mu$T. Since the conventional contrast was 3.3 \% under a static magnetic field of 5.0 $\mu$T when using the neutral density filter instead of the analyzer, the crossed polarizers method obtained about 25 times better contrast than the conventional method did. In addition, the linewidth obtained with the crossed polarizers method was 1.15 kHz, which was about twice as narrow as the conventional method's value of 2.15 kHz. This result means that the resonance with the crossed polarizers method has not only higher contrast but also a narrower linewidth.

\subsection{Contrast as a function of the relative angle $\theta$}

\begin{figure}[!b]
\centering
\includegraphics[scale=0.5]{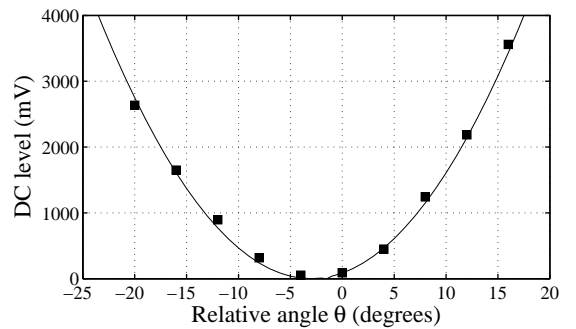}
\caption{DC level as a function of the relative angle $\theta$: The square dot is the measured data, and the solid line is the fitting curve calculated from Eq. (\ref{eq:i_dc}). The magnetic field was set to 93 $\mu$T.}
\label{fig:DClevel trans angle}
\end{figure}

\begin{figure}[!b]
\centering
\includegraphics[scale=0.5]{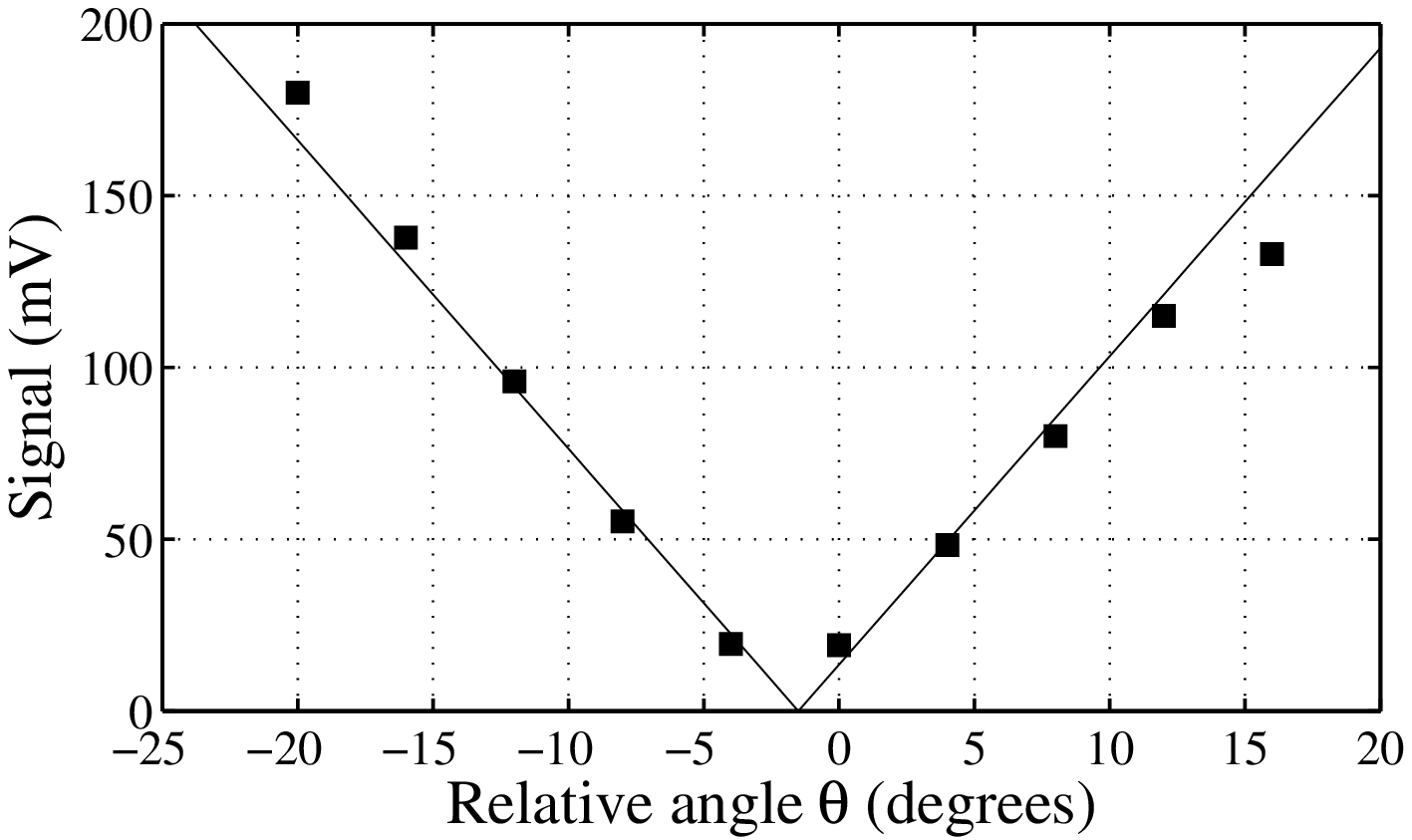}
\caption{Signal as a function of the relative angle $\theta$: The square dot is the measured data, and the solid line is the fitting curve calculated from Eq. (\ref{eq:i_s}). The magnetic field was set to 93 $\mu$T.}
\label{fig:Signal trans angle}
\end{figure}

\begin{figure}[!b]
\centering
\includegraphics[scale=0.5]{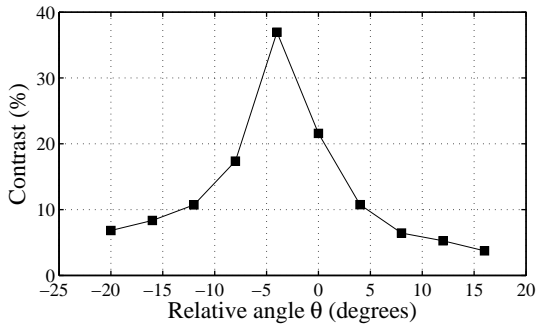}
\caption{Contrast as a function of the relative angle $\theta$.}
\label{fig:contrast trans angle}
\end{figure}

Figure \ref{fig:DClevel trans angle} shows the DC level as a function of the relative angle $\theta$. The square dot is the measured data, and the solid line is the fitting curve calculated from Eq. (\ref{eq:i_dc}). The relative angle $\theta$ giving the minimum DC level is shifted from 0$^\circ$. This shift is caused by misalignment between the scale of the polarizer's mount and the transmission axis of polarizers. The extinction ratio of the polarizers was estimated to be about 40 dB. 

Figure \ref{fig:Signal trans angle} shows the signal of the resonance as a function of the relative angle $\theta$. The square dot is the measured data, and the solid line is the fitting curve calculated from Eq. (\ref{eq:i_s}). The calculated curve is in good agreement with the experimental data. By comparing the signal and DC level, it can be seen that the behavior of the signal was different from that of the DC level. Therefore, this is proof that the polarization of the wavelength components contributing to CPT is rotated.

Figure \ref{fig:contrast trans angle} shows the contrast as a function of the relative angle $\theta$ from the measurements. Since the signal has a high value despite the DC level reduction, the contrast significantly increased near 0$^\circ$. A resonance contrast over 10\% was obtained in the range from -15 to 5$^\circ$. 

\subsection{Characteristics as a function of magnetic field}

\begin{figure}[!b]
\centering
\includegraphics[scale=0.5]{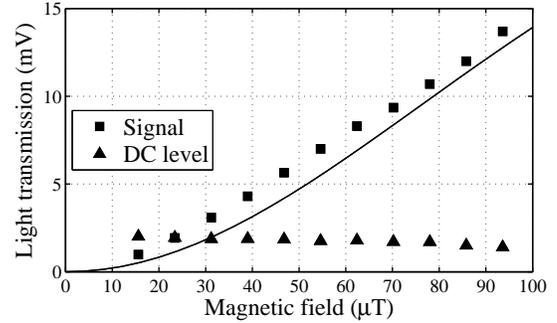}
\caption{Signal and DC level as a function of magnetic field: The square and triangle are the measured signal and DC level, respectively. The solid line is a fit of Eq. (\ref{eq:i_pp}): resonance linewidth $\gamma$ = 1.08 kHz.}
\label{fig:SD magnetic field}
\end{figure}

\begin{figure}[!b]
\centering
\includegraphics[scale=0.5]{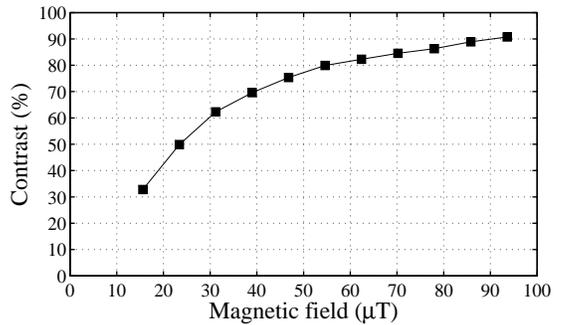}
\caption{Contrast as a function of magnetic field estimated from Fig. \ref{fig:SD magnetic field}.}
\label{fig:contrast magnetic field}
\end{figure}

\begin{figure}[!b]
\centering
\includegraphics[scale=0.5]{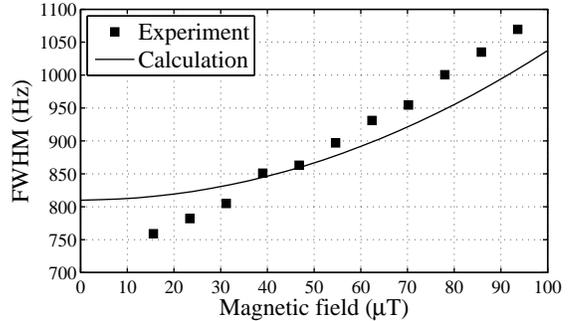}
\caption{Linewidth as a function of magnetic field: The solid line is data calculated from Eq. (\ref{eq:gamma_cp}): resonance linewidth $\gamma$ = 1.08 kHz.}
\label{fig:linewidth magnetic field}
\end{figure}

\begin{figure}[!b]
\centering
\includegraphics[scale=0.5]{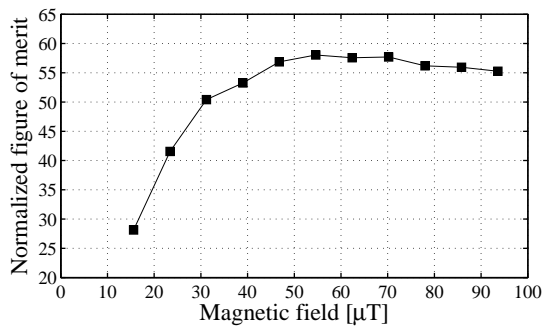}
\caption{Figure of merit as a function of magnetic field. The figure of merit is normalized by that of the conventional resonance.}
\label{fig:figure of merit magnetic field}
\end{figure}

Figure \ref{fig:SD magnetic field} shows the signal and DC level as a function of magnetic field. The relative angle $\theta$ was optimized in order to maximize contrast. In weak magnetic fields ($<$ 15 $ \mu$T), the signal was so small that we could not observe CPT resonance, and the maximum magnetic field (93 $\mu$T) was limited by the current source output supplying the Helmholz coil. The signal linearly increased with increasing magnetic field. On the other hand, the DC level was constant regardless of the change in the magnetic field. The results are evidence that the Faraday rotation affected only the wavelength components contributing to CPT. The solid line is a fitting curve of Eq. (\ref{eq:i_pp}). The experimental signal has the same tendency as the theoretical curve.

Figure \ref{fig:contrast magnetic field} shows the contrast estimated from the measurement results in Fig. \ref{fig:SD magnetic field}. The contrast increased with increasing magnetic field. We assume that nearly 100 \% contrast can be obtained in a larger magnetic field. The DC level is independent of the magnetic field, and this indicates that the resonance contrast has reached a peak value. From Eq. (\ref{eq:i_pp}), it is estimated that the maximum contrast of 94.0\% can be obtained at 224 $\mu$T.

The linewidth as a function of the magnetic field is shown in Fig.\ref{fig:linewidth magnetic field}. The linewidth broadened with increasing magnetic field and was approximately proportional to it. The linewidth obtained with the crossed polarizers method is less than 2.15 kHz of the linewidth with the conventional excitation method in this range of magnetic field. At a magnetic field of 15 $\mu$T, the narrowest linewidth obtained was 760 Hz, which is about three times narrower than the conventional one. The solid line in Fig. \ref{fig:linewidth magnetic field} is calculated data based on Eq. (\ref{eq:gamma_cp}). The difference between the measurement and calculation is no more than 20 \%.

Figure \ref{fig:figure of merit magnetic field} shows the figure of merit (FoM) as a function of magnetic field. Since short-term stability is determined by the contrast and linewidth from Eq. (\ref{eq:short}), the FoM is defined as follows.

\begin{equation}
\mbox{FoM} := \frac{f_0}{\Delta f}\cdot \mbox{Contrast}
\label{eq:fhfs}
\end{equation}

In small magnetic fields ($<$ 40 $\mu$T), the FoM increased because the increase in contrast was dominant. However, in large magnetic fields ($>$ 60 $\mu$T), the FoM decreased with broadening linewidth. This shows that the FoM of the CPT resonance has a peak value with respect to the magnetic field. In this experiment, the maximum value of FoM was obtained at 55 $\mu$T, and this value is 58 times better than the conventional one. 

\section{Conclusion}
We developed a new method based on crossed polarizers for observing high-contrast CPT resonance. Firstly, we calculated the Faraday rotation in CPT under a lin$\parallel$lin light field by using two pairs of $\Lambda$ system models. The calculated results indicated that the resonance amplitude has a peak value with respect to the magnetic field, and the resonance linewidth increases with increasing magnetic field. The minimum linewidth obtained with the crossed polarizers method is 36.8\% that of the conventional method in the absence of the magnetic field. Secondly, we measured the resonance characteristics with crossed polarizers method using a $^{133}$Cs gas cell and the D$_1$-line VCSEL. It was shown that the background signal level of the photodetector is suppressed by the crossed polarizers method and a high contrast resonance can be obtained. As the relative angle $\theta$ and magnetic field were optimized, the observed resonance had a contrast of 88.4\% and linewidth of 1.15 kHz. The measurement data was in good agreement with the theoretical data and the difference between experiment and theory was no more than 20\%. Finally, we investigated the optimal conditions for enhancing short-term stability. The figure of merit has a peak value with respect to magnetic field. By optimizing the relative angle $\theta$ of the analyzer and magnetic field, the figure of merit was 58 times better than the conventional one. Since a high contrast and narrow linewidth resonance can be obtained with such a simple optical system, the crossed polarizers method is an attractive means of enhancing the frequency stability of CPT atomic clocks.

\section*{Acknowledgments}
The authors are grateful to Ricoh Company, Ltd. for providing us with the Cs-D$_1$ VCSEL.


\begin{IEEEbiography}[{\includegraphics[width=1in,height=1.25in,clip,keepaspectratio,clip,keepaspectratio]{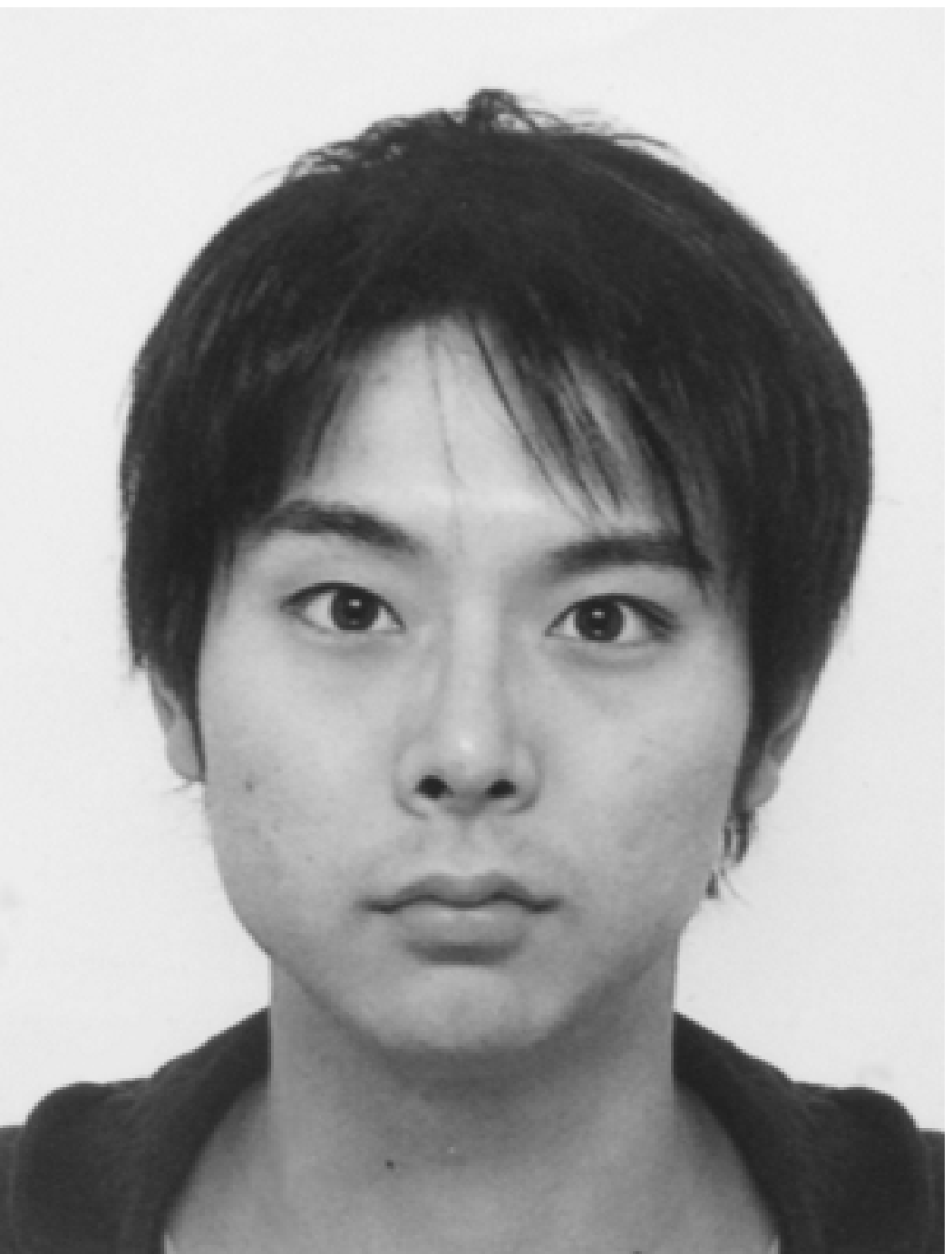}}]{Yuichiro Yano} was born in Tokyo, Japan, in 1987. He received M. E. degree in electronics from Tokyo Metropolitan University, Hachioji, Japan, in 2012.
\end{IEEEbiography}

\begin{IEEEbiography}[{\includegraphics[width=1in,height=1.25in,clip,keepaspectratio,clip,keepaspectratio]{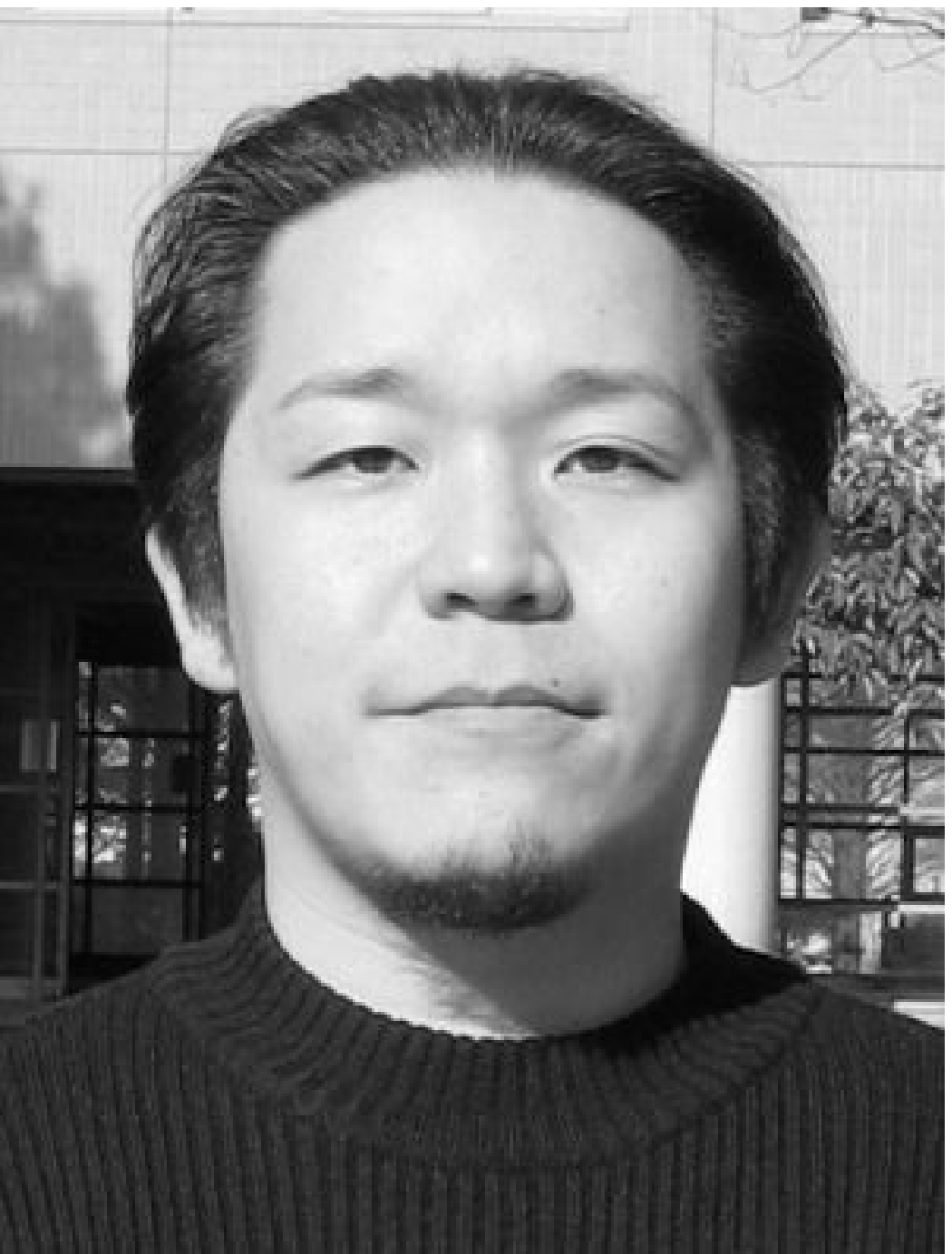}}]{Shigeyoshi Goka} received the M. E. degree from Tokyo Metropolitan University, Tokyo, Japan in 1997, and the Ph. D. degree in engineering from same university in 2005. He is an associate professor in the Department of Electrical \& Electronic Engineering, Graduate School of Tokyo Metropolitan University, where he is engaged in research on the miniature atomic clocks, the magnetometer, and the design of mesa shaped quartz resonators. Dr. Goka is a member of the Institute of Electronics, Information, and Communication Engineers (IEICE) of Japan, and the Institute of Electrical Engineers of Japan (IEEJ). He is a manager of the Technical Committee of Precise Frequency, the Technical Committee of Electro-Mechanical Devices, and the Technical Committee of Modeling and Simulations of the IEEJ.
\end{IEEEbiography}


\begin{thebibliography}{1}
\bibitem{Knappe}  
S.~Knappe, V.~Shah, P.~D.~Schwindt, L.~Hollberg, and J.~Kitching, L.~Liew and J.~Moreland, "A microfabricated atomic clock", {\it Appl. Phys. Lett.}, vol. 85, no. 9, pp.1460-1462, Aug. 2004.
\bibitem{Vig}  
J.~Vig, "Military Applications of High Accuracy Frequency Standards and Clocks", {\it IEEE Trans. Ultrason. Ferroelectr. Freq. Control}, vol. 40, no. 5, pp.522-527, Sep. 1993.
\bibitem{Lutwak}  
R.~Lutwak, D.~Emmons, T.~English and W.~Riley, "The chip-scale atomic clock-recent development progress", in {\it Proc. 35th Annu. Precise Time and Time Interval Systems and Applications Meeting}, pp.467-478, 2004.
\bibitem{Powerbroadening}  
S.~Knappe, R.~Wynands, J.~Kitching, H.~G.~Robinson, and L.~Hollberg, "Characterization of coherent population-trapping resonances as atomic frequency references", {\it J. Opt. Soc. Am. B}, vol. 18, no. 11,  p.1545-1553, Nov. 2001.

\bibitem{Pulsed CPT}  
T.~Zanon, S.~Guerandel, E.~de~Clercq, D.~Holleville, N.~Dimarcq, and A.~Clairong, "High Contrast Ramsey Fringes with Coherent-Population-Trapping Pulses in a Double Lambda Atomic System", {\it Phys. Rev. Lett.}, vol. 94, art. no. 193002, May. 2005.

\bibitem{Push-pull}  
Y.-Y.~Jau, E.~Miron, A.~B.~Post, N.~N.~Kuzma, and W.~Happer, "Push-Pull Optical Pumping of Pure Superposition States", {\it Phys. Rev. Lett}, vol. 93, art. no. 160802, Oct. 2004.


\bibitem{Magneto}
S.~Pradhan, R.~Behera and A.~K.~Das, "Polarization rotation under two-photon Raman resonance for magnetometry",
{\it Appl. Phys. Lett.}, vol. 100 art. no. 1783502, 2012.

\bibitem{POP}
J.~Lin, J.~Deng, Y.~Ma, H.~He and Y.~Wang, "Detection of ultrahigh resonance contrast in vapor-cell atomic clocks", {\it Optics Lett.}, vol. 37, pp. 5036-5038, 2012.


\bibitem{Budker}  
D.~Budker, D.~F.~Kimball and D.~P.~Demille, "Atomic physics: an exploration through problems and solutions", Oxford University Press, 2004.

\bibitem{Zibrov}
S.~A.~Zibrov, I.~Novikova, D.~F.~Phillips, R.~L.~Walsworth, A.~S.~Zibrov, V.~L.~Velichansky, A.~V.~Taichenachev and V.~I.~Yudin, "Coherent-population-trapping resonances with linearly polarized light for all-optical miniature atomic clocks", {\it Phys. Rev. A}, vol. 81, art. no. 013833, 2010.

\bibitem{Kitching}
J.~Kitching, H.~G.~Robinson, L.~Hollberg, S.~Knappe and R.~Wynands, "Optical-pumping noise in laser-pumped, all-optical microwave frequency references", {\it J. Opt. Soc. Am. B}, vol. 18, no. 11, pp.1676-1683, Nov. 2001.

\bibitem{Dick}
B.~Dick, "High-contrast polarization spectroscopy of photochemically burned spectral holes in amorphous solids: Potential for fast optical storage" {\it Chemical Phys. Lett.}, vol. 143, no. 2, pp.186-192, Jan. 1988.

\bibitem{Budker2}
D.~Budker, W.~Gawlik, D.~F.~Kimball, S.~M.~Rochester, V.~V.~Yashchuk, and A.~Weis, "Resonant nonlinear magneto-optical effects in atoms" {\it Rev. Mod. Phys.}, vol. 74, pp. 1153-1201, 2002.

\bibitem{contrast}
K.~Watabe, T.~Ikegami, A.~Takamizawa, S.~Yanagimachi, S.~Ohshima, and S.~Knappe, "High-contrast dark resonances with linearly polarized light on the D$_1$ line of alkali atoms with large nuclear spin", {\it Applied Optics}, vol. 48, no. 6, pp.1098-1103, Feb. 2009.



\end{thebibliography}
\end{document}